\RequirePackage{lineno}
\documentclass[twocolumn,showpacs,aps,prd]{revtex4}

\usepackage{graphicx}
\usepackage{dcolumn}
\usepackage{amsmath}
\usepackage{epsfig}
\usepackage{subfigure}
\usepackage{epstopdf} 
\usepackage{array}
\newcolumntype{C}[1]{>{\centering\arraybackslash}p{#1}}
\usepackage{color}

\begin{document}
\normalsize

\parskip=5pt plus 1pt minus 1pt

\title{\boldmath Observation of the doubly radiative decay $\eta^{\prime}\to \gamma\gamma\pi^0$}

\author{
{\small
M.~Ablikim$^{1}$, M.~N.~Achasov$^{9,d}$, S. ~Ahmed$^{14}$, X.~C.~Ai$^{1}$, O.~Albayrak$^{5}$, M.~Albrecht$^{4}$, D.~J.~Ambrose$^{45}$, A.~Amoroso$^{50A,50C}$, F.~F.~An$^{1}$, Q.~An$^{47,38}$
% ...
\\
 \vspace{0.2cm}
 (BESIII Collaboration)\\
 \vspace{0.2cm} {\it
$^{1}$ Institute of High Energy Physics, Beijing 100049, People's Republic of China\\
% ...
$^{4}$ Bochum Ruhr-University, D-44780 Bochum, Germany\\
$^{5}$ Carnegie Mellon University, Pittsburgh, Pennsylvania 15213, USA\\
% ...
$^{9}$ G.I. Budker Institute of Nuclear Physics SB RAS (BINP), Novosibirsk 630090, Russia\\
% ...
$^{14}$ Helmholtz Institute Mainz, Johann-Joachim-Becher-Weg 45, D-55099 Mainz, Germany\\
% ...
$^{38}$ State Key Laboratory of Particle Detection and Electronics, Beijing 100049, Hefei 230026, People's Republic of China\\
% ...
$^{45}$ University of Rochester, Rochester, New York 14627, USA\\
% ...
$^{47}$ University of Science and Technology of China, Hefei 230026, People's Republic of China\\
% ...
$^{50}$ (A)University of Turin, I-10125, Turin, Italy; (B)University of Eastern Piedmont, I-15121, Alessandria, Italy; (C)INFN, I-10125, Turin, Italy\\
% ...
$^{d}$ Also at the Novosibirsk State University, Novosibirsk, 630090, Russia\\
% ...
}} \vspace{0.4cm} }

\begin{abstract}
Based on a sample of $1.31$ billion $J/\psi$
events collected with the BESIII detector, we report the study of the doubly radiative decay $\eta^\prime\to \gamma\gamma\pi^0$ for the first time, where the $\eta^\prime$ meson is produced via the $J/\psi\to \gamma\eta^\prime$ decay.
The branching fraction of $\eta^\prime\to \gamma\gamma\pi^0$ inclusive decay is measured to be ${\cal B}(\eta^\prime\to \gamma\gamma\pi^0)_{\text{Incl.}}$
= $(3.20\pm0.07\mbox{(stat)}\pm0.23\mbox{(sys)})\times 10^{-3}$, while the branching fractions  of the dominant process $\eta^\prime\rightarrow\gamma\omega$ and the nonresonant component are
determined to be ${\cal B}(\eta^\prime\to \gamma\omega)\times {\cal B}(\omega\to \gamma\pi^0) = (23.7 \pm1.4\mbox{(stat)}\pm1.8\mbox{(sys)})\times 10^{-4}$ and ${\cal B}(\eta^\prime\to \gamma\gamma\pi^0)_{\text{NR}} = (6.16\pm0.64\mbox{(stat)} \pm0.67\mbox{(sys)})\times 10^{-4}$, respectively. In addition, the $M^2_{\gamma\gamma}$-dependent partial widths of the inclusive decay are also presented.
\end{abstract}

\pacs{13.40.Gp, 13.40.Hq, 13.20.Jf, 14.40.Be}

\maketitle

\normalsize

\section{Introduction}

The $\eta^\prime$ meson provides a unique stage for understanding the distinct symmetry-breaking mechanisms present in low-energy
quantum chromodynamics (QCD)~\cite{qcd-sym1,qcd-sym2,qcd-sym3,bes3-eta,bes3-eta2} and its decays play an important role in exploring
the effective theory of QCD at low energy~\cite{XPhT}. Recently, the doubly radiative decay $\eta^\prime\to \gamma\gamma\pi^0$ was
studied in the frameworks of the linear $\sigma$ model (L$\sigma$M) and the vector meson dominance (VMD) model~\cite{VMD_LsigmaM_0,VMD_LsigmaM}.
It has been demonstrated that the contributions from the VMD are dominant. Experimentally, only an upper limit of the nonresonant
branching fraction of ${\cal B}(\eta^\prime\to \gamma\gamma\pi^0)_{\text{NR}}<8\times 10^{-4}$ at the 90\% confidence level has been
determined by the GAMS-2000 experiment~\cite{GAMS_3}.

In this article, we report the first measurement of the branching fraction of the inclusive $\eta^\prime\to \gamma\gamma\pi^0$ decay
and the determination of the $M^2_{\gamma\gamma}$ dependent partial widths, where $M_{\gamma\gamma}$ is the invariant mass of the
two radiative photons. The inclusive decay is defined as the $\eta^\prime$ decay into the final state $\gamma\gamma\pi^0$ including
all possible intermediate contributions from the $\rho$ and $\omega$ mesons below the $\eta^\prime$ mass threshold and the nonresonant
contribution from the excited vector meson above the $\eta^\prime$ mass threshold. Since the contribution from mesons above the $\eta^\prime$ threshold actually derives from the low-mass tail and looks like a contact term, we call
this contribution 'nonresonant'. The branching fraction for the nonresonant $\eta^\prime\to \gamma\gamma\pi^0$ decay is obtained from a fit to the $\gamma\pi^0$ invariant
mass distribution by excluding the coherent contributions from the $\rho$ and $\omega$ intermediate states. The measurement of the
$M^2_{\gamma\gamma}$ dependent partial widths will provide direct inputs to the theoretical calculations on the transition form factors
of $\eta^\prime\to \gamma\gamma\pi^0$ and improve the theoretical understanding of the $\eta^\prime$ decay mechanisms.

\section{Experimental Details}

The source of $\eta^\prime$ mesons is the radiative $J/\psi\to \gamma\eta^\prime$ decay in a sample of $1.31\times 10^{9}$ $J/\psi$
events~\cite{NJpsi09, NJpsi} collected by the BESIII detector. Details on the features and capabilities of the BESIII detector can
be found in Ref.~\cite{BEPCII}.

The response of the BESIII detector is modeled with a Monte Carlo (MC) simulation based on {\sc geant4}~\cite{geant1}. The program
{\sc evtgen}~\cite{evtgen} is used to generate a $J/\psi\to \gamma\eta^\prime$ MC sample with an angular distribution of
$1 + \cos^2\theta_\gamma$, where $\theta_\gamma$ is the angle of the radiative photon relative to the positron beam direction in
the $J/\psi$ rest frame. The decays $\eta^\prime\to \gamma\omega(\rho)$, $\omega(\rho)\to \gamma\pi^0$ are generated using the helicity
amplitude formalism. For the nonresonant $\eta^\prime\to \gamma\gamma\pi^0$ decay, the VMD model~\cite{VMD_LsigmaM_0,VMD_LsigmaM} is used
to generate the MC sample with $\rho(1450)$ or $\omega(1650)$ exchange. Inclusive $J/\psi$ decays are generated with {\sc kkmc}~\cite{kkmc}
generator; the known $J/\psi$ decay modes are generated by {\sc evtgen}~\cite{evtgen} with branching fractions setting at Particle
Data Group (PDG) world average values~\cite{PDG14}; the remaining unknown decays are generated with {\sc lundcharm}~\cite{lund}.

\section{Event Selection and Background Estimation}

Electromagnetic showers are reconstructed from clusters of energy deposits in the electromagnetic calorimeter (EMC). The energy deposited
in nearby time-of-light (TOF) counters is included to improve the reconstruction efficiency and energy resolution. The photon candidate
showers must have a minimum energy of 25~MeV in the barrel region ($|\cos\theta|<0.80$) or 50~MeV in the end cap region ($0.86<|\cos\theta|<0.92$).
Showers in the region between the barrel and the end caps are poorly measured and excluded from the analysis. In this analysis, only
the events without charged particles are subjected to further analysis. The average event vertex of each run is assumed as the origin
for the selected candidates. To select $J/\psi\to \gamma\eta^\prime$, $\eta^\prime\to \gamma\gamma\pi^0$ $(\pi^0\to \gamma\gamma)$
signal events, only the events with exactly five photon candidates are selected.

\begin{figure}[tb]
\centering
  \includegraphics[width=6.5cm]{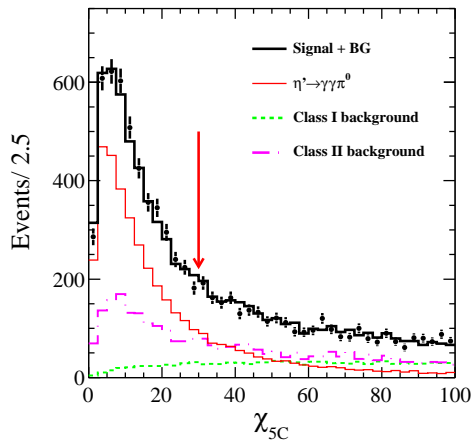}
  \caption{Distribution of the $\chi^{2}_{5C}$ of the 5C kinematic fit for the inclusive $\eta^{\prime}$ decay. Dots with error bars are
  data; the heavy (black) solid-curve is the sum of signal and expected backgrounds from MC simulations; the light (red) solid-curves
  is signal components which are normalized to the fitted yields; the (green) dotted-curve is the class I background; and the (pink)
  dot-dashed-curve is the class II background.}
\label{m2gam_bf5c}
\end{figure}

\begin{figure}
  \centering
  \includegraphics[width=6.5cm]{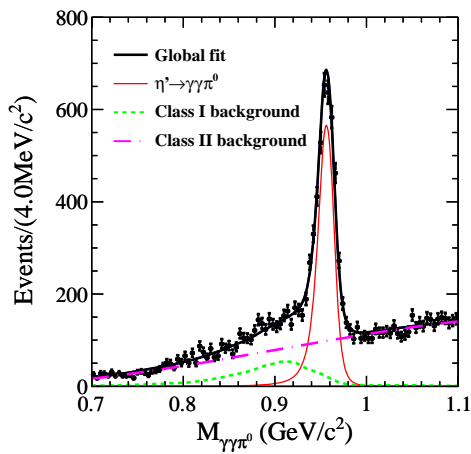}
  \caption{Results of the fit to $M_{\gamma\gamma\pi^0}$ for the selected inclusive $\eta^\prime\to \gamma\gamma\pi^0$ signal events.
  The (black) dots with error bars are the data.}
  \label{etafit_R}
  \vspace{0.0cm}
\end{figure}

To improve resolution and reduce background, a five-constraint kinematic (5C) fit imposing energy-momentum conservation and a $\pi^0$
mass constraint is performed to the $\gamma\gamma\gamma\pi^0$ hypothesis, where the $\pi^0$ candidate is reconstructed with a pair
of photons. For events with more than one $\pi^0$ candidate, the combination with the smallest $\chi^{2}_{5\mbox{c}}$ is selected.
Only events with $\chi^{2}_{5\mbox{c}}<30$ are retained. The $\chi^{2}_{5C}$ distribution is shown in Fig.~\ref{m2gam_bf5c} with
events in the $\eta^{\prime}$ signal region of  $|M_{\gamma\gamma\pi^{0}} - M_{\eta^{\prime}}|<25$~MeV ($M_{\eta^{\prime}}$ is the
$\eta^\prime$ nominal mass from PDG~\cite{PDG14}). In order to suppress the multi-$\pi^0$ backgrounds and remove the miscombined
$\pi^0$ candidates, an event is vetoed if any two of five selected photons (except for the combination for the $\pi^0$ candidate)
satisfies $|M_{\gamma\gamma} - M_{\pi^0}|<18$~MeV/c$^2$, where $M_{\pi^0}$ is the $\pi^0$ nominal mass. After the application
of the above requirements, the most energetic photon is taken as the primary photon from the $J/\psi$ decay, and the remaining two
photons and the $\pi^0$ are used to reconstruct the $\eta^\prime$ candidates. Figure~\ref{etafit_R} shows the $\gamma\gamma\pi^0$
invariant mass spectrum.

%

Detailed MC studies indicate that no peaking background remains after all the selection criteria. The sources of backgrounds are
divided into two classes. Background events of class I are from $J/\psi\to \gamma\eta^\prime$ with $\eta^\prime$ decaying into final
states other than the signal final states. These background events accumulate near the lower side of the $\eta^\prime$ signal region
and are mainly from $\eta^\prime\to \pi^0\pi^0\eta$ ($\eta\to \gamma\gamma$), $\eta^\prime\to 3\pi^0$ and $\eta^\prime\to \gamma\gamma$,
as shown as the (green) dotted curve in Fig.~\ref{etafit_R}. Background events in class II are mainly from $J/\psi$ decays to final
states without $\eta^\prime$, such as $J/\psi\to \gamma\pi^0\pi^0$ and $J/\psi\to \omega\eta$ ($\omega\to \gamma\pi^0$, $\eta\to \gamma\gamma$)
decays, which contribute a smooth distribution under the $\eta^\prime$ signal region as displayed as the (pink) dot-dashed curve
in Fig.~\ref{etafit_R}.
\begin{table*}[htbp]
  \begin{center}
  \caption{
    Observed $\eta^\prime$ signal yields ($N^{\eta^\prime}$) and detection efficiencies ($\epsilon$) for inclusive $\eta^\prime\to \gamma\gamma\pi^0$, $\eta^\prime\to \gamma\omega$($\omega\to \gamma\pi^0$), and the nonresonant $\eta^\prime\to \gamma\gamma\pi^0$ decays. The measured
    branching fractions$^c$ in this work, comparison of values from the PDG~\cite{PDG14} and theoretical predictions are listed. The
    first errors are statistical and the second ones are systematic.}
  \setlength{\extrarowheight}{1.0ex}
  \renewcommand{\arraystretch}{0.9}
  \vspace{0.2cm}
  \begin{tabular}{ccccc}
    \hline\hline
                       &$\eta^\prime\to \gamma\gamma\pi^0$ (Inclusive)    &$\eta^\prime\to \gamma\omega, \omega\to \gamma\pi^0$
      &$\eta^\prime\to \gamma\gamma\pi^0$ (Nonresonant) \\  \hline
    $N^{\eta^\prime}$  &$3435\pm76\pm244$                                 &$2340\pm141\pm180$
      &$655\pm68\pm71$  \\
    $\epsilon$         &16.1\%                                            &14.8\%
      &15.9\%  \\ \

    ${\mathcal B}~(10^{-4})$               &$32.0\pm0.7\pm2.3$     &$23.7\pm1.4\pm1.8^{a}$
      &$6.16\pm0.64\pm0.67$  \\
    ${\mathcal B}_{\text{PDG}}~(10^{-4})$  &--                                              &$21.7\pm1.3^{b}$
      &$<8$  \\
    Predictions $(10^{-4})$                &57~\cite{VMD_LsigmaM_0},65~\cite{VMD_LsigmaM}   &--
      &--  \\
    \hline\hline
  \end{tabular}
  \vspace{-0.2cm}
  \label{tab:br}
  \begin{flushleft}
  \footnotesize{$^{a}$ The product branching fraction $\mathcal{B}(\eta^\prime \to \gamma\omega)\cdot \mathcal{B}(\omega\to \gamma\pi^0)$.}
  \footnotesize{$^{b}$ The product branching fraction $\mathcal{B}(\eta^\prime \to \gamma\omega)\cdot \mathcal{B}(\omega\to \gamma\pi^0)$
  from PDG~\cite{PDG14}.}
  \footnotesize{$^{c}$ The product branching fraction $\mathcal{B}(\eta^\prime \to \gamma\rho^0)\cdot \mathcal{B}(\rho^0\to \gamma\pi^0)$
  is determined to be $(1.92\pm0.16(\text{stat}))\times 10^{-4}$ using the fitted yield in Fig.~\ref{Inter_fit}, which is in agreement
  with the PDG value of $(1.75\pm0.23)\times 10^{-4}$~\cite{PDG14}.}
  \end{flushleft}
  \end{center}
\end{table*}
%

\section{Signal Yields and Branching Fractions}

A fit to the $\gamma\gamma\pi^0$ invariant mass distribution is performed to determine the inclusive $\eta^\prime\to \gamma\gamma\pi^0$
signal yield. The probability density function (PDF) for the signal component is represented by the signal MC shape, which is obtained
from the signal MC sample generated with an incoherent mixture of $\rho$, $\omega$ and the nonresonant components according to the
fractions obtained in this analysis. Both the shape and the yield for the class I background are fixed to the MC simulations and
their expected intensities. The shape for the class II background is described by a third-order Chebychev polynomial, and the corresponding
yield and PDF parameters are left free in the fit to data. The fit range is 0.70$-$1.10~GeV/c$^2$. Figure~\ref{etafit_R} shows
the results of the fit. The fit quality assessed with the binned distribution is $\chi^2/\text{n.d.f}=108/95=1.14$. The signal yield
and the MC-determined signal efficiency for the inclusive $\eta^\prime$ decay are summarized in Table~\ref{tab:br}.

\begin{figure}
  \centering
  \includegraphics[width=6.5cm]{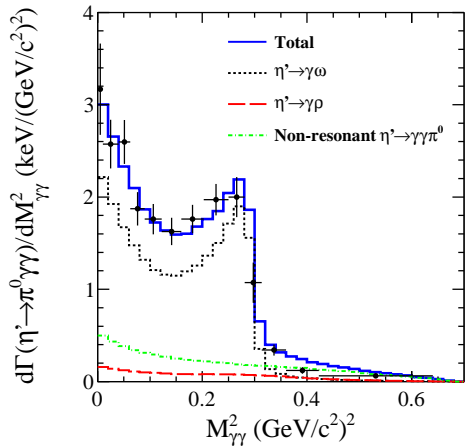}
  \caption{
    Partial width (in keV) versus $M^2_{\gamma\gamma}$ for the inclusive $\eta^\prime\to \gamma\gamma\pi^0$ decay. The error includes
    the statistic and systematic uncertainties. The (blue) histogram is the sum of an incoherent mixture of $\rho$-$\omega$ and the
    nonresonant components from MC simulations; the (back) dotted-curves is $\omega$-contribution; the (red) dot-dashed-curve is
    the $\rho$-contribution; and the (green) dashed-curve is the nonresonant contribution. All the components are normalized using
    the yields obtained in Fig.~\ref{Inter_fit}.}
  \label{Form_factor}
\end{figure}

\begin{figure}
  \centering
  \includegraphics[width=6.5cm]{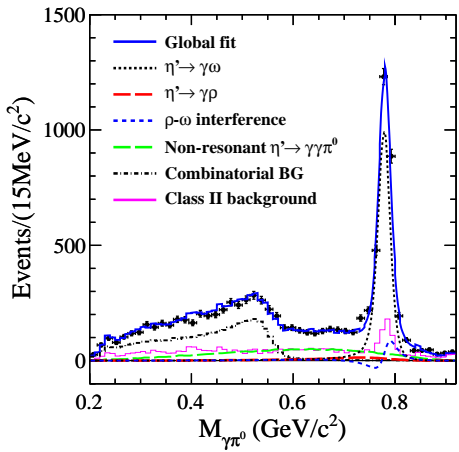}
    \caption{
      Distribution of the invariant mass $M_{\gamma\pi^0}$ and fit results in the $\eta^\prime$ mass region. The points with error
      bars are data; the (black) dotted-curve is from the $\omega$-contribution; the (red) long dashed-curve is from the $\rho$-contribution;
      the (blue) short dashed-curve is the contribution of $\rho$-$\omega$ interference; the (green) long dashed curve is the nonresonance;
      the (pink) histogram is from the class II background; the (black) short dot-dashed curve is the combinatorial backgrounds of
      $\eta^\prime\to \gamma\omega$, $\gamma\rho$. The (blue) solid line shows the total fit function.}
    \label{Inter_fit}
\end{figure}

In this analysis, the partial widths can be obtained by studying the efficiency-corrected signal yields for each given $M^2_{\gamma\gamma}$
bin $i$ for the inclusive $\eta^\prime \to \gamma\gamma\pi^0$ decay. The resolution in $M^2_{\gamma\gamma}$ is found to be about
$5\times10^2$~(MeV/c$^2)^2$ from the MC simulation, which is much smaller than $1.0\times 10^4$~(MeV/c$^2)^2$, a statistically reasonable
bin width, and hence no unfolding is necessary. The $\eta^\prime$ signal yield in each $M^2_{\gamma\gamma}$ bin is obtained by performing bin-by-bin
fits to the $\gamma\gamma\pi^0$ invariant mass distributions using the fit procedure described above. Thus the background-subtracted, efficiency-corrected
signal yield can be used to obtain the partial width for each given $M^2_{\gamma\gamma}$ interval, where the PDG value is used for
the total width of the $\eta^{\prime}$ meson~\cite{PDG14}. The results for $d\Gamma(\eta^\prime\to \gamma\gamma\pi^0)/dM^2_{\gamma\gamma}$
in each $M^2_{\gamma\gamma}$ interval are listed in Table~\ref{tab:BR-FF} and depicted in Fig.~\ref{Form_factor}, where the contributions from each component obtained from the MC simulations are normalized with the yields by fitting to $M_{\gamma\pi^0}$ as displayed in Fig.~\ref{Inter_fit}.

%
\begin{table*}[htbp]
\centering
  \caption{Results for $d\Gamma(\eta^\prime\to \gamma\gamma\pi^0)/dM^2_{\gamma\gamma}$ (in units of keV/(GeV/c$^2)^2$)
  for thirteen intervals of $M^2_{\gamma\gamma}$. The first uncertainties  are statistical and the second systematic.}
  \footnotesize
  \setlength{\extrarowheight}{1.0ex}
  \renewcommand{\arraystretch}{1.0}
  \vspace{0.2cm}
  \begin{tabular}{c|c|c|c|c|c}
   \hline\hline
   $M^2_{\gamma\gamma}$ ((GeV/c$^2)^2)$  &$[0.0, 0.01]$        &$[0.01, 0.04]$       &$[0.04, 0.06]$       &$[0.06, 0.09]$      &$[0.09, 0.12]$ \\
   $d\Gamma(\eta^\prime\to \gamma\gamma\pi^0)/M^2_{\gamma\gamma}$  &$3.17\pm0.44\pm0.24$ &$2.57\pm0.18\pm0.19$ &$2.60\pm0.15\pm0.18$ &$1.87\pm0.12\pm0.14$&$1.76\pm0.11\pm0.13$ \\
   \hline\hline
   $M^2_{\gamma\gamma}$ ((GeV/c$^2)^2)$  &$[0.12, 0.16]$       &$[0.16, 0.20]$       &$[0.20, 0.25]$       &$[0.25, 0.28]$      &$[0.28, 0.31]$ \\
   $d\Gamma(\eta^\prime\to \gamma\gamma\pi^0)/M^2_{\gamma\gamma}$          &$1.63\pm0.10\pm0.12$ &$1.76\pm0.09\pm0.13$ &$1.97\pm0.10\pm0.14$ &$2.00\pm0.17\pm0.15$&$1.07\pm0.20\pm0.08$ \\
   \hline\hline
   $M^2_{\gamma\gamma}$ ((GeV/c$^2)^2)$  &$[0.31, 0.36]$       &$[0.36, 0.42]$       &$[0.42, 0.64]$       &                    &  \\
   $d\Gamma(\eta^\prime\to \gamma\gamma\pi^0)/M^2_{\gamma\gamma}$          &$0.34\pm0.06\pm0.03$ &$0.12\pm0.03\pm0.01$ &$0.06\pm0.01\pm0.01$ &                    &  \\
   \hline\hline
  \end{tabular}
\vspace{0.1cm}
\label{tab:BR-FF}
\end{table*}
%

Assuming that the inclusive decay $\eta^\prime\to \gamma\gamma\pi^0$ can be attributed to the vector mesons $\rho$ and $\omega$ and
the nonresonant contribution, we apply a fit to the $\gamma\pi^0$ invariant mass to determine the branching fraction for the nonresonant
$\eta^\prime\to \gamma\gamma\pi^0$ decay using the $\eta^\prime$ signal events with $|M_{\gamma\gamma\pi^0} - m_{\eta^\prime}|<25$~MeV/c$^{2}$.
In the fit, the $\rho$-$\omega$ interference is considered, but possible interference between the $\omega$ ($\rho$) and the nonresonant
process is neglected. To validate our fit, we also determine the product branching fraction for the decay chain $\eta^\prime\to \gamma\omega$,
$\omega\to \gamma\pi^0$. Figure~\ref{Inter_fit} shows the $M_{\gamma\pi^0}$ distribution. Since the doubly radiative photons are
indistinguishable, two entries are filled into the histogram for each event. For the PDF of the coherent $\omega$ and $\rho$ produced in
$\eta^\prime\to \gamma\gamma\pi^0$, we use $[\varepsilon(M_{\gamma\pi^0})\times E^3_{\gamma^{\eta^\prime}}\times E^3_{\gamma^{\omega(\rho)}}\times |\text{BW}_{\omega}(M_{\gamma\pi^0}) + \alpha e^{i\theta}\text{BW}_{\rho}(M_{\gamma\pi^0})|^2\times \text{B}^2_{\eta^\prime}\times \text{B}^2_{\omega(\rho)}]\otimes \text{G}(0, \sigma)$, where $\varepsilon(M_{\gamma\pi^{0}})$ is the detection efficiency determined
by the MC simulations; $E_{\gamma^{\eta^\prime(\omega/\rho)}}$ is the energy of the transition photon in the rest frame of $\eta^\prime$
($\omega/\rho$); $\text{BW}_{\omega}(M_{\gamma\pi^0})$ is a relativistic Breit-Wigner (BW) function, and $\text{BW}_{\rho}(M_{\gamma\pi^0})$
is a relativistic BW function with mass-dependent width~\cite{GS}. The masses and widths of the $\rho$ and $\omega$ meson are fixed
to their PDG values~\cite{PDG14}. $\text{B}^2_{\eta^\prime(\omega/\rho)}$ is the Blatt-Weisskopf centrifugal barrier factor for the
$\eta^\prime$($\omega/\rho$) decay vertex with radius $R=0.75$ fm~\cite{BR-factor1,BR-factor2}, and $\text{B}^2_{\eta^\prime(\omega/\rho)}$
is used to damp the divergent tail due to the factor $E^3_{\gamma^{\eta^\prime(\omega/\rho)}}$.  The Gaussian function $\text{G}(0, \sigma)$
is used to parameterize the detector resolution. The combinatorial background is produced by the combination of the $\pi^0$ and the
photon from the $\eta^\prime$ meson, and its PDF is described with a fixed shape from the MC simulation. The ratio of yields between
the combinatorial backgrounds and the coherent sum of $\rho$-$\omega$ signals is fixed from the MC simulations. The shape of the
nonresonant signal $\eta^\prime\to \gamma\gamma\pi^0$ is determined from the MC simulation, and its yield is determined in the fit.
The background from the class I as discussed above is fixed to the shape and yield of the MC simulation.  Finally, the shape from
the class II background is obtained from the $\eta^\prime$ mass sidebands (738$-$788 and 1008$-$1058~MeV/c$^{2}$), and its normalization
is fixed in the fit. The $M_{\gamma\pi^0}$ mass range used in the fit is 0.20$-$0.92~GeV/c$^2$. In the fit, the interference phase $\theta$ between the $\rho$- and $\omega$-components is allowed. Due to the low statistics of the $\rho$ meson contribution, we fix the ratio $\alpha$ of $\rho$ and
$\omega$ intensities to the value for the ratio of ${\cal B}(\eta^\prime \to \gamma\rho)\cdot {\cal B}(\rho\to \gamma\pi^0)$ and
${\cal B}(\eta^\prime \to \gamma\omega)\cdot{\cal B}(\omega\to \gamma\pi^0)$ from the PDG~\cite{PDG14}. Figure~\ref{Inter_fit} shows
the results. The yields for the vector mesons $\rho$, $\omega$ and their interference are determined to be $(183\pm15)$, $(2340\pm141)$,
and $(174\pm92)$, respectively. The signal yields and efficiencies as well as the corresponding branching fractions for the
$\eta^\prime\to \gamma\omega(\omega\to \gamma\pi^0)$ and nonresonant decays are summarized in Table~\ref{tab:br}.

\section{Systematic Uncertainties}

The systematic uncertainties on the branching fraction measurements are summarized in Table~\ref{tab:error}. The uncertainty due
to the photon reconstruction is determined to be 1\% per photon as described in Ref.~\cite{number}.
The uncertainties associated with the other selection criteria, kinematic fit
with $\chi^2_{5C}<30$, the number of photons equal to 5 and $\pi^0$ veto ($|M_{\gamma\gamma} - M_{\pi^0}|>18$~MeV/c$^2$)
are studied with the control sample $J/\psi\to \gamma\eta^{\prime}$, $\eta^{\prime}\to \gamma\omega$, $\omega\to \gamma\pi^{0}$ decay, respectively.
The systematic error in each of the applied selection criteria is numerically estimated from the ratio of the number of events with and without the corresponding requirement.
The corresponding resulting efficiency differences between data and MC (2.7\%, 0.5\%, and 1.9\% , respectively) are taken to be representative of the corresponding systematic uncertainties.

In the fit for the inclusive $\eta^\prime$ decay, the signal shape is fixed to the MC simulation. The uncertainty due to the signal shape is considered by convolving a Gaussian
function to account for the difference in the mass resolution between data and MC simulation.
In the fit
to the $\gamma\pi^{0}$ distribution, alternative fits with the mass
resolution left free in the fit and the radius $R$ in the barrier
factor changed from 0.75~fm to 0.35~fm are performed, and the changes
of the signal yields are taken as the uncertainty due to the signal shape.

In the fit to the $M_{\gamma\gamma\pi^{0}}$ distribution, the signal
shape is described with an incoherent sum of contributions from
processes involving $\rho$ and $\omega$
and nonresonant processes obtained from MC simulation, where the nonresonant process is modeled with the VMD model. A fit with an alternative signal model for the different components, \emph{i.e.}
a coherent sum for the $\rho$-, $\omega$-components and a uniform angular distribution in phase space (PHSP) for the nonresonant
process, is performed. The resultant changes in the branching fractions
are taken as the uncertainty related to the signal model. An alternate fit to the $M_{\gamma\pi^{0}}$ distribution
is performed, where the PDF of the nonresonant decay is extracted
from the PHSP MC sample. The changes in the measured branching
fractions are considered to be the uncertainty arising from the signal model.

In the fit to the $M_{\gamma\pi^{0}}$ distribution, the uncertainty due to the fixed relative $\rho$ intensity is evaluated by
changing its expectation by one standard deviation.
An alternative fit in which the ratio of yields between combinatorial
backgrounds and the coherent sum of $\rho-\omega$ signals is changed by one standard deviation from the MC simulation is performed, and the change observed
in the signal yield is assigned as the uncertainty.
A series of fits using different fit ranges is performed and the maximum change of the branching fraction is taken as a systematic
uncertainty.

The uncertainty due to the class I background is estimated by varying the numbers of expected background events by one standard
deviation according to the errors on the branching fraction values in PDG~\cite{PDG14}. The uncertainty due to the class II background
is evaluated by changing the order of the Chebychev polynomial from 3
to 4 for the fit to the $\eta^{\prime}$ inclusive decay, and varying
the ranges of $\eta^{\prime}$ sidebands for the fit to the $\gamma\pi^{0}$ invariant mass distribution, respectively.

The number of $J/\psi$ events is $N_{J/\psi} = (1310.6\pm 10.5)\times 10^{6}$~\cite{NJpsi09,NJpsi}, corresponding to an uncertainty
of 0.8\%. The branching fractions for the $J/\psi\to \gamma\eta^\prime$ and $\pi^0\to \gamma\gamma$ decays are taken from the PDG~\cite{PDG14},
and the corresponding uncertainties are taken as a systematic uncertainty. The total systematic errors are 7.1\%, 7.7\%, 10.8\% for
the inclusive decay, $\omega$ contribution and nonresonant decay, respectively, as summarized in Table~\ref{tab:error}.

\begin{table}[htbp]
  \centering
  \caption{Summary of relative systematic uncertainties ($\%$) for the branching fraction measurements. Here $\eta^\prime_{\text{Incl.}}$, $\eta^{\prime}_\omega$ and $\eta^{\prime}_{\text{NR}}$ represent the inclusive $\eta^\prime\to \gamma\gamma\pi^0$, $\eta^\prime\to \gamma\omega
  (\omega\to \gamma\pi^0$) and nonresonant decays, respectively.}
  \setlength{\extrarowheight}{1.0ex}
  \renewcommand{\arraystretch}{0.8}
  \vspace{0.2cm}
  \begin{tabular}{C{4.0cm}|*{3}{C{1.cm}}}
  \hline\hline
                                 &$\eta^\prime_{\text{Incl.}}$  &$\eta^{\prime}_\omega$  &$\eta^{\prime}_{\text{NR}}$ \\  \hline
   Photon detection              &5.0                           &5.0                     &5.0  \\
   5C kinematic fit              &2.7                           &2.7                     &2.7  \\
   Number of photons             &0.5                           &0.5                     &0.5  \\
   $\pi^{0}$ veto                &1.9                           &1.9                     &1.9 \\

   Signal shape                  &0.5                           &1.5                     &2.3  \\
   Signal model                  &1.7                           &1.0                     &4.3  \\

   $\rho$ relative intensity     &--                            &1.3                     &4.9  \\
   Combinatorial backgrounds     &--                            &1.3                     &0.8  \\

   Fit range                     &0.8                           &1.6                     &2.1  \\

   Class I background            &0.1                            &0.2                    &0.6  \\
   Class II background           &0.3                            &1.8                    &4.2  \\

   Cited branching fractions     &3.1                           &3.1                     &3.1  \\

   Number of $J/\psi$ events     &0.8                           &0.8                     &0.8  \\\hline

   Total systematic error        &7.1                           &7.7                     &10.8\\
  \hline\hline
  \end{tabular}
  \vspace{0.0cm}
  \label{tab:error}
\end{table}
%

\section{Summary}

In summary, with a sample of $1.31\times 10^{9}$ $J/\psi$ events collected with the BESIII detector, the doubly radiative decay
$\eta^{\prime}\to \gamma\gamma\pi^{0}$ has been studied. The branching fraction of the inclusive decay is measured for the first
time to be ${\cal B}(\eta^{\prime}\to \gamma\gamma\pi^{0})_{\text{Incl.}} = (3.20\pm0.07\mbox{(stat)}\pm0.23\mbox{(sys)})\times 10^{-3}$.
The $M^{2}_{\gamma\gamma}$ dependent
partial decay widths are also determined. In addition, the branching fraction for the nonresonant decay is determined to be
${\cal B}(\eta^{\prime}\to \gamma\gamma\pi^{0})_{\text{NR}}$ = $(6.16\pm0.64\mbox{(stat)}\pm0.67\mbox{(sys)})\times 10^{-4}$, which
agrees with the upper limit measured by the GAMS-2000 experiment~\cite{GAMS_3}. As a validation of the fit, the product branching
fraction with the omega intermediate state involved is obtained to be ${\cal B}(\eta^{\prime}\to \gamma\omega)\cdot{\cal B}(\omega\to \gamma\pi^{0})$
= $(2.37\pm0.14\mbox{(stat)} \pm0.18\mbox{(sys)})\times 10^{-3}$, which is consistent with the PDG value~\cite{PDG14}. These results
are useful to test QCD calculations on the transition form factor, and provide valuable inputs to the theoretical understanding of
the light meson decay mechanisms.

\section*{Acknowledgments}

The BESIII Collaboration thanks the staff of BEPCII and the IHEP computing center for their strong support. This work is supported
in part by National Key Basic Research Program of China under Contract No. 2015CB856700; National Natural Science Foundation of China
(NSFC) under Contracts No. 11125525, No. 11235011, No. 11322544, No. 11335008, No. 11335009, No. 11425524, No. 11505111, No. 11635010,
No. 11675184; the Chinese Academy of Sciences (CAS) Large-Scale Scientific Facility Program; the CAS Center for Excellence in Particle
Physics (CCEPP); the Collaborative Innovation Center for Particles and Interactions (CICPI); Joint Large-Scale Scientific Facility
Funds of the NSFC and CAS under Contracts No. U1232201, No. U1332201, No. U1532257, No. U1532258; CAS under Contracts No. KJCX2-YWN29,
No. KJCX2-YW-N45; 100 Talents Program of CAS; National 1000 Talents Program of China; INPAC and Shanghai Key Laboratory for Particle
Physics and Cosmology; German Research Foundation DFG under Contracts No. Collaborative Research Center CRC 1044, FOR 2359; Istituto
Nazionale di Fisica Nucleare, Italy; Koninklijke Nederlandse Akademie van Wetenschappen (KNAW) under Contract No. 530-4CDP03; Ministry
of Development of Turkey under Contract No. DPT2006K-120470; The Swedish Research Council; U.S. Department of Energy under Contracts
No. DE-FG02-05ER41374, No. DE-SC-0010118, No. DE-SC-0010504, No. DE-SC-0012069; U.S. National Science Foundation; University of Groningen
(RuG) and the Helmholtzzentrum fuer Schwerionenforschung GmbH (GSI), Darmstadt; WCU Program of National Research Foundation of Korea
under Contract No. R32-2008-000-10155-0.



\begin{thebibliography}{99}
%
\bibitem{qcd-sym1}
J. Steinberger, Phys. Rev. {\bf 76}, 1180 (1949); S. L. Adler, Phys. Rev. {\bf 177}, 2426 (1969); J. S. Bell and R. Jackiw,
Nuovo Cim. {\bf A 60}, 47 (1969); W. A. Bardeen, Phys. Rev. {\bf 184}, 1848 (1969).

%
\bibitem{qcd-sym2}
J. Wess and B. Zumino, Phys. Lett. {\bf B 37}, 95 (1971).

%
\bibitem{qcd-sym3}
E. Witten, Nucl. Phys. {\bf B223}, 422 (1983).

\bibitem{bes3-eta} H. B. Li, J. Phys. {\bf G 36}, 085009 (2009).

\bibitem{bes3-eta2} A. Kupsc, Int. J. Mod. Phys. {\bf E 18}, 1255 (2009).
%
\bibitem{XPhT}
J. Gasser and H. Leutwyler, Nucl. Phys. {\bf B250}, 465 (1985); H. Neufeld and H. Rupertsberger, Z. Phys. {\bf C 68}, 91 (1995).

\bibitem{VMD_LsigmaM_0}
R. Jora, Nucl. Phys. Proc. Suppl. 207-208, 224 (2010).

%
\bibitem{VMD_LsigmaM}
R. Escribano, PoS QNP 2012, 079 (2012).

%
\bibitem{GAMS_3}
D. Alde {\it et al.} (GAMS-2000 Collaboration), Z. Phys. {\bf C 36}, 603 (1987).

\bibitem{NJpsi09}
M. Ablikim {\it et al.} (BESIII Collaboration), Chin. Phys. {\bf C 36}, 915 (2012).

%
\bibitem{NJpsi}
M.~Ablikim {\it et al.} [BESIII Collaboration],
  	Chin. Phys. \textbf{C 41}, 013001 (2017).
%
%
\bibitem{BEPCII}
M. Ablikim {\it et al.} (BES Collaboration), Nucl. Instrum. Methods Phys. Res., Sect. {\bf A 614}, 345 (2010).

%
\bibitem{geant1}
S. Agostinelli {\it et al.} (GEANT4 Collaboration), Nucl. Instrum. Methods Phys. Res., Sect.  {\bf A 506}, 250
(2003).

%
\bibitem{evtgen}
D.~J.~Lange, Nucl.\ Instrum.\ Meth. {\bf A 462}, 1 (2001);
R.~G.~Ping, Chin.\ Phys. {\bf C 32}, 599 (2008).


\bibitem{kkmc}
S. Jadach, B. F. L. Ward and Z. Was, Comput. Phys. Commun. {\bf 130}, 260 (2000);  Phys. Rev. {\bf D 63}, 113009 (2001).

%
\bibitem{PDG14}
C. Patrignani  {\it et al.} (Particle Data Group), Chin. Phys. {\bf C 40}, 100001 (2016).

%
\bibitem{lund}
J. C. Chen, G. S. Huang, X. R. Qi, D. H. Zhang, and Y. S. Zhu, Phys. Rev. {\bf D 62}, 034003 (2000).


%
\bibitem{GS}
J. P. Lees {\it et al.} (BaBar Collaboration), Phys. Rev. {\bf D 88}, 032013 (2013).

%
\bibitem{BR-factor1}
S. U. Chung, Phys. Rev. {\bf D 48}, 1225 (1993).

%
\bibitem{BR-factor2}
F. Hippel, C. Quigg, Phys. Rev. {\bf D 5}, 624 (1972).

%
\bibitem{number}
M. Ablikim {\it et al.} (BESIII Collaboration), Phys. Rev. {\bf D 83}, 012003 (2011).

\end{thebibliography}
\end{document}